\documentclass[]{spie}  %

\usepackage{amsmath,amsfonts,amssymb}
\usepackage{graphicx}
\usepackage{aas_macros}
\usepackage[colorlinks=true, allcolors=blue]{hyperref}

\title{The Black Hole Explorer: Back End Electronics}

\author[a]{Ranjani Srinivasan}
\author[a,b]{Jonathan Weintroub}
\author[h]{Rick Raffanti}
\author[c]{Bryan Bilyeu}
\author[a]{Thomas Gauron}
\author[a]{John Test}
\author[a]{Elliot Richards}
\author[d]{Manuel Fernandez}
\author[a]{Mark Freeman}
\author[a]{Peter Cheimets}
\author[f]{Mauricio Gendelman}
\author[a]{Kari Haworth}
\author[a]{Janice Houston}
\author[a,b]{Michael D. Johnson}
\author[d]{Emilia Mamani}
\author[g]{Daniel  Marrone}
\author[e]{Ariel L. Pola}
\author[c]{Jade Wang}

\affil[a]{Center for Astrophysics $|$ Harvard \& Smithsonian, 60 Garden St., Cambridge, MA 02138, USA}
\affil[b]{Black Hole Initiative at Harvard University, 20 Garden Street, Cambridge, MA 02138, USA}
\affil[c]{MIT Lincoln Laboratory, Lexington, MA 02421, USA}
\affil[d]{National University of Córdoba, Av. Haya de la Torre, Córdoba, Argentina}
\affil[e]{Celero Communication Inc., Lund 22221, Sweden}
\affil[f]{University of Buenos Aires, Paseo Colón 850, Ciudad de Buenos Aires, Argentina}
\affil[g]{University of Arizona, Steward Observatory, Tucson, AZ 85719, USA}
\affil[h]{Techne Instruments, 4920 Telegraph Ave, Unit G.  Oakland, CA 94609, USA}

\authorinfo{Further author information: (Send correspondence to R.S.)\\R.S.: E-mail: ranjani.srinivasan@cfa.harvard.edu}

\begin{document} 
\maketitle

\begin{abstract}

This paper describes specification and early design of back end signal processing subsystems for the Black Hole Explorer (BHEX) Very Long Baseline Interferometry (VLBI) space telescope. The``back end'' consists of two subsystems.  First, the block downconveter (BDC) is a heterodyne system that performs a frequency translation of the analog signal from IF to baseband and amplifies and filters it for digitization. Second, the digital back end (DBE) samples the analog signal with an analog-to-digital converters (ADC) and digitally processes the data stream formatting them to the VLBI ``VDIF'' standard and converting to Ethernet packets for 100\,gigabit-per-second (Gb/s) Ethernet transport to the optical downlink system \cite{BHEX_Wang_2024}.  Both the BDC and the DBE for BHEX support eight channels of 4.096~GHz bandwidth each, for a total processed bandwidth of 32.768\,GHz. The BHEX back end benefits from mature terrestrial back end heritage, described in some detail.  The BHEX back end itself is in the early stages of design, with requirements, interface specifications, and component trade studies well advanced. The aim is to build a prototype using terrestrial grade parts which are available in functionally identical space grade equivalents, and to use this prototype to advance the back end Technology Readiness Level (TRL) preparing for a Small Explorer (SMEX) proposal in 2025.

\end{abstract}


\section{INTRODUCTION}
\label{sec:intro}  %

The Black Hole Explorer (BHEX) \cite{BHEX_Johnson_2024} is a $\sim$1\,mm Very Long Baseline Interferometry (VLBI) station in a mid-Earth orbit (MEO). BHEX will discover, measure, and make images of the bright and narrow ``photon ring'' that is predicted to exist around black holes. The photon ring corresponds to light that has circled the black hole---sometimes multiple times---before escaping, and is a fine structure underlying the Event Horizon Telescope (EHT) black hole ring images of M87* {\cite{EHT2019a}, and SgrA* \cite{SgrAEHTCI}.   Making an image of the photon ring is not possible on terrestrial EHT baselines which are limited to one earth diameter.  BHEX extends the EHT to space allowing $\sim 2 \times$ the baseline length, and enabling the necessary resolution. This paper presents a description and initial design of BHEX signal processing elements following the dual receivers \cite{BHEX_Tong_2024} and prior to the optical data downlink system.\cite{Wang_2023},\cite{BHEX_Wang_2024}

 Section \ref{sec:system} briefly describes the BHEX electronics, and section \ref{sec:previous_work} describes progenitor terrestrial  back end systems. The BHEX back end work includes trade studies, requirements and interface development, and firmware design, and is described in section \ref{sec:DBE}. The first objective is a prototype, built with terrestrial equivalents of space grade parts, used for functional and environmental testing and TRL maturation, to prepare for a SMEX mission proposal in 2025.

This paper \cite{BHEX_Srinivasan_2024} is a part of the series of papers describing the BHEX mission\cite{BHEX_Johnson_2024, BHEX_Akiyama_2024,  BHEX_Marrone_2024, BHEX_Peretz_2024, BHEX_Lupsasca_2024, BHEX_Galison_2024, BHEX_Issaoun_2024, BHEX_Kawashima_2024, BHEX_Tomio_2024, BHEX_Wang_2024, BHEX_Sridharan_2024, BHEX_Rana_2024, BHEX_Tong_2024} submitted to the SPIE Astronomical Telescopes $+$ Instrumentation 2024 conference. Specific relevant papers are cross-cited where appropriate in the narrative.

\section{BHEX System Overview}

\label{sec:system}

The major subsystems of the BHEX science instrument are a 3.5~meter antenna.  The antenna is coupled to a coherent dual band, dual-polarization receiver, then to the back end described in this paper which in turn feeds dat to a high throughput optical downlink \cite{BHEX_Wang_2024}.  We provide a brief description of the BHEX subsystems other than the back end as context for the latter's detailed description in section \ref{sec:DBE}.  For more detail of the entire instrument please see the relevant companion paper in this proceedings \cite{BHEX_Marrone_2024}.

\subsection{Dish Antenna}
As the space-based component of a space-ground interferometer, the sensitivity of each space-ground baseline (formed by correlating the electric fields received simultaneously at BHEX and a ground-based observatory) is proportional to the geometric mean of the space and ground antenna collecting areas.   This allows BHEX to be implemented with a 3.5\,m dish \cite{BHEX_Sridharan_2024}  suitable for launch in available fairings.

\subsection{Dual Band Receivers}

 The receivers \cite{BHEX_Tong_2024}  observe simultaneously in two bands.  The ``low'' band is centered at 86\,GHz.  The ``high'' band has a center frequency widely tunable between 220 and 325\,GHZ.   Each receiver has wide sidebands of 8.192\,GHz,  and both polarizations are detected, so each  band has a total width of 16.384\,GHz---32.768\,GHz across both receivers.  The wide bandwidth improves signal-to-noise-ratio (SNR) according to the radiometer equation, but results in large data payloads which are a challenge to downlink.  \footnote{The receiver for the high band is a so-called ``double sideband'' or DSB architecture, meaning that two sidebands (upper and lower) on the sky are overlaid onto a single receiver IF output.  Thus the total dual band sky frequency coverage is in fact 32.768\,GHz.  The upper and lower sidebands are separated at the terrestrial VLBI correlator after data downlink.  The DSB approach aims to reduce downlink payload for this very wide sky bandwidth which is invisible to the back end electronics, and ``sees'' the unseparated 16.384\,GHz bandwidth. The low band receiver is a single sideband (USB) system with dual polarization.}.   The receivers operate at cryogenic temperatures, the high band requiring a cold stage of 4\,K. Space qualified cryogenics are specialized systems discussed in a companion paper in this proceedings \cite{BHEX_Rana_2024}.

The effect of the atmosphere on the phase of the wavefront received at ground stations increases with frequency resulting in significant coherence losses at high observing frequencies. A major motivation for BHEX and its associated ground array to observe simultaneously in two bands, is to enable Frequency Phase Transfer (FPT) \cite{BHEX_Issaoun_2024}. FPT transfers phase calibration from the low band to the high band. The low band has greater sensitivity and is less affected by the atmosphere, however low band VLBI even on the BHEX array long base baseline has insufficient resolution to resolve the photon ring. The primary purpose of the low band is to provide a phase calibration of the high band whose short wavelength results in fine angular resolution sufficient to resolve the photon ring. Also in the case of SgrA* the high band sees more clearly to the core of the source, and the image is less blurred by scattering in the interstellar medium foreground. 

\subsection{Frequency reference}
An ultra-stable oscillator provides the master frequency reference, to ensure coherence of the BHEX space station on timescales sufficient to find correlation detections, or ``fringes''.   Both quartz oscillators and a new breed of compact iodine atomic clocks are in consideration for this critical subsystem function.

\subsection{Optical Data Downlink}
The TBIRD (TeraByte Infrared Delivery) is a small cubesat mission in low-earth orbit (LEO) and recently (2022, 2023)\cite{Wang_2023}  demonstrated up to 200 Gb/s downlink capacity.   This is far more  data than available from typical radio-frequency links or prior space  optical communications.  The Optical Data Downlink subsystem under development for BHEX is the evolution of the TBIRD system and is discussed in detail in companion papers in this proceedings \cite{BHEX_Marrone_2024}, \cite{BHEX_Wang_2024}.   The optical downlink subsystem lies immediately downstream of the BHEX DBE, which is one of the two subsystems described in detail in the present paper.

\section{Progenitor Terrestrial Back End Subsystems}
\label{sec:previous_work}

The development of the BHEX back end subsystems benefit from considerable terrestrial experience developing EHT back ends subsystems.  Our Digital Signal Processing (DSP) team at the Smithsonian Astrophysical Observatory (SAO) has considerable experience in the application of wideband samplers (or analog-to-digital converters, ADC) and FPGAs in the design of signal processing systems for ground-based submillimeter interferometry. We have designed and commissioned the terrestrial back ends for the Event Horizon Telescope (EHT)\cite{Vertatschitsch_2015} and correlator-beamformer systems for the Submillimeter Array (SMA)  \cite{SWARM}.

Back ends for the next generation Event Horizon Telescope (ngEHT) \cite{ngEHT_Doeleman} capable of conditioning and simultaneously digitizing  4 x 8.192\,GHz of instantaneous analog bandwidth are in advanced development.  The ngEHT back end comprises an analog signal processing component (ngBDC) and a digitizer and digital signal processor board (ngDBE).  The ngDBE development includes a considerable library of firmware for ADC interfacing, signal processing, and an Ethernet stack. The ngEHT design including both hardware and firmware elements provides considerable experience and heritage benefiting the design of the BHEX back ends.  

The installed and commissioned EHT back ends are discussed in \ref{sec:r2dbe}, and the next generation back end subsystems in sections \ref{sec:ngbdc} and \ref{sec:ngDBE}.  The ngDBE FPGA firmware is discussed in \ref{sec:ngdbefw} noting that the considerable effort in this firmware can readily be leveraged and ported to the BHEX DBE.

\subsection{ROACH2 Digital Back End (R2DBE)}
\label{sec:r2dbe}
The current EHT back end  was developed in 2014, and is called the ROACH2 Digital Back End \cite[R2DBE]{Vertatschitsch_2015}. 
The R2DBE samples 2 x 2.048\,GHz of instantaneous analog bandwidth. It uses a Xilinx Virtex 6 FPGA as its digital signal processing engine to re-quantize 8-bit samples to 2-bit resolution, pack the data in VDIF format, and transmit them over two 10\,Gb/s Ethernet connections to the data recorder. Accurate timing and synchronization is achieved by referencing to a hydrogen maser.  The ngDBE has a matching block downconverter (BDC) developed in Europe at NOVA Laboratories. The R2DBE was successfully deployed to all single-dish EHT telescopes that observed in 2017, resulting in the first EHT images of M87*\cite{EHT2019a} and SgrA*\cite{EHT2021a}.

\subsection{next generation Block Downconverter (ngBDC)}
\label{sec:ngbdc}

The ngBDC performs a frequency translation of the receiver output from IF to baseband and conditions it for digitization by the ngDBE. For simultaneous tri frequency, dual polarization and dual sideband operations, twelve such signal processing chains are required.  The prototype was manufactured by Xmicrowave, Inc, using drop-in modules representative of the final ngBDC circuit board. The prototype was characterized and performance was found to be well within the specifications. The prototype was converted to a production custom printed circuit board in a fieldable enclosure. Shown in Figure~\ref{fig:BDC_pictures_Sweep} are layout of the production unit (with the two custom filters seen) on the top left panel, a photo of the production version housed in an EMI enclosure (right panel) and a VNA sweep of a representative signal chain (bottom left panel), the performance across the passband looks excellent.
 
 \begin{figure*}[h]
    \centering
    \includegraphics[width=\textwidth]{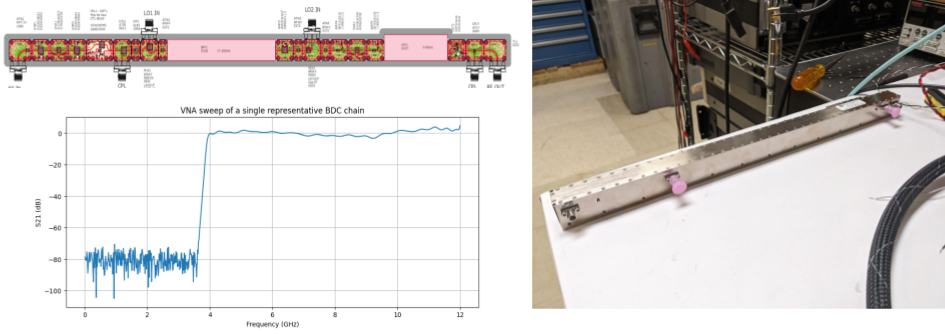}\vspace{0cm}
    \caption{(left)(top) The PCB layout of a single  production level BDC signal chain. The two "hollow" regions are the custom bandpass and Nyquist filters.(right) A photo of a single BDC chain enclosed in an EMI housing. (left)(bottom) A VNA sweep of a representative signal chain over 8.192 GHz.  As seen, the passband response looks excellent.  }
\label{fig:BDC_pictures_Sweep} 
\end{figure*}

The design incorporates the following features:
\begin{enumerate}
\item  A sophisticated downconversion scheme employing an “up-down” conversion method to eliminate mixing of input and output signals.
\item Test points using input and output stage couplers to help with debugging.
\item Two custom filters to provide image rejection and to provide a sharp Nyquist cutoff to prevent aliasing.
\item Second stage mixing using spectral inversion.
\item An input stage closed loop servo control using a Variable Voltage Attenuator (VVA) to prevent any compression arising from higher input power levels. 
\item An output stage open loop Automatic Gain Control (AGC) using a VVA to load the digitizer optimally.
\end{enumerate}
 The field deployable units are modular in design with respect to frequency which will allow for ease of installation and future expansion.

\subsection{next generation Digital Back End (ngDBE)}
\label{sec:ngDBE}

\begin{figure*}[h]
    \centering
    \includegraphics[width=\textwidth]{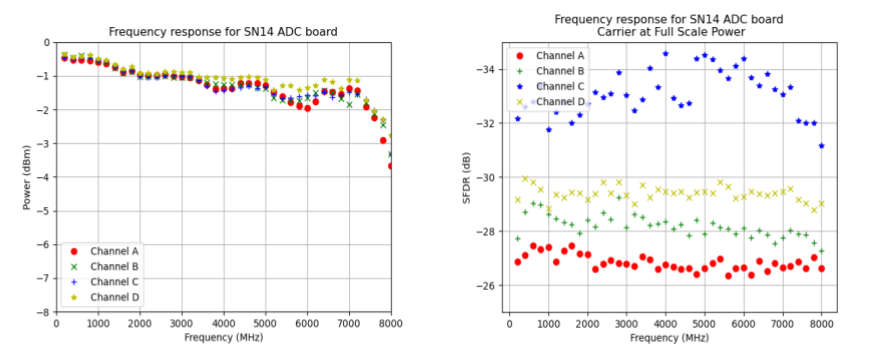}\vspace{0cm}
    \caption{The ngDBE ADCs are driven at full scale for both these sine wave the measurements. (left) Plot of the frequency response of the 4 ADC channels. Data is collected at intervals of 200~MHz as a tone sweeps in frequency across the band. A standing wave pattern is seen across the entire band for all the 4 channels, albeit severely mitigated up to about 7.5 GHz. The overall slope across the DC to 8~GHz band is approximately 3 dB. (right) The spurious free dynamic range (SFDR) is also measured at intervals of 200 MHz.  Each point is an average of 10 measurements. The 2nd harmonic is the strongest tone apart from the fundamental frequency, and there is a very weak frequency dependence seen across the band.}
\label{fig:FreqResp_SFDR} 
\end{figure*}

\begin{figure*}[h]
    \centering
    \includegraphics[width=\textwidth]{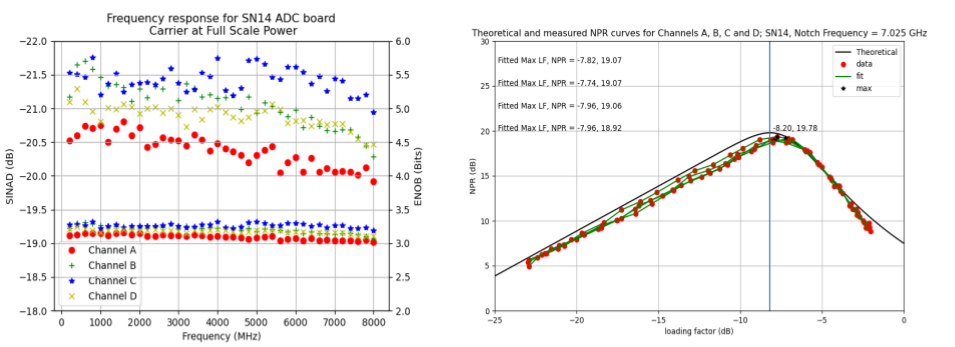}\vspace{0cm}
    \caption{(left) The ngEHT DBE signal-to-noise-and-distortion ration (SINAD), or, equivalently,  effective number of bits (ENOB) of the 4-bit converer are measured at intervals of 200~MHz starting at 200 MHz up to 8~GHz. At each frequency, the ADC is driven at full scale, and each data point is an average of 10 measurements. The left axis depicts the SINAD and the right axis depicts the corresponding ENOB. As expected, the 4 channels have somewhat different performance profiles. The ENOB is above 3-bits across the first Nyquist zone.
(right) The red points are  measured noise power ratio (NPR) values averaged over 100 FFT spectra to drive down the noise. The black solid line is the theoretical value. The green curve is a fitted spline, and is used to compute the maximum of the NPR curve to find the corresponding loading factor which is the optimal loading for the 4 bit ADC. The blue vertical line is the maximum for the theoretical curve. The four measured and fitted lines correspond to the four ADC channels on a single board and show good concurrence with the theoretical curves and with each other. }
\label{fig:ENOB_NPR} 
\end{figure*}

\begin{figure*}[h]
    \centering
    \includegraphics[width=\textwidth]{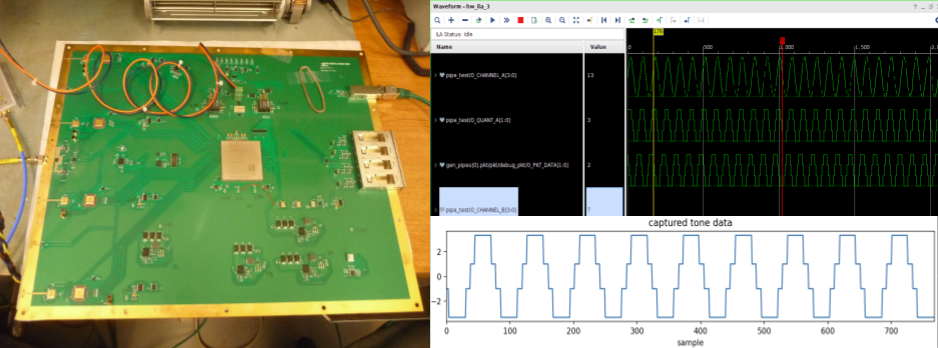}\vspace{0cm}
    \caption{(left) A photo of the first revision of the integrated (ADC and FPGA chips on the same substrate) ngDBE circuit board with the  components installed.   Using this pre-production prototype, received a few weeks before this paper was submitted,  we have run test firmware developed on the evaluation board.  The right panel shows measured results of this test. The  top three panels showing digitized, requantized and packetized data from an ILA on the FPGA.The bottom panel showing a reconstructed sine wave from data captured on an external capture device. }
\label{fig:DBE} 
\end{figure*}

Each ngDBE board accepts four 8.192 GHz baseband inputs. The Nyquist criterion for sampling requires sampling the data at 16.384 Gsps. Each ngDBE board is outfitted with four 4-bit high speed wideband analog to digital converters from Adsantec and a VU13P, an Ultrascale+ Virtex FPGA from AMD Xilinx and other ancilliary components. Two 64 Gbps streams of packetized 2-bit data are sent out to be shipped over 100 GbE interfaces to recorders. 

The high speed analog to digital converters were fully characterized, and several relevant figures of merit are measured and compared to theoretical expectations. The frequency response of the ADC and Spurious Free Dynamic Range (SFDR) measurements are plotted in Figure ~\ref{fig:FreqResp_SFDR}. The frequency response is very flat with only a 3~dB slope across the 8 GHz passband. Similarly the SFDR is well within expected levels for such a wideband ADC. The other two key figures of merit namely the Effective Number of Bits (ENOB), which is derived from SIgnal to Noise And Distortion (SINAD) and the Noise Power ratio (NPR) are shown in Figure~\ref{fig:ENOB_NPR} on the left and right panels respectively. The ENOB stays above 3 bits and above across the entire band. We consider this performance quite acceptable noting the that the recorded VLBI data are quantized  to 2-bits. 

\subsection{ngDBE Firmware}
\label{sec:ngdbefw}
The platform for firmware development is the prototype ngDBE system, which uses a quad custom sampler PCB board   interfaced via an FMC+ (Vita 57.4) connector to an evaluation board.  The VCU128 is an AMD (formerly Xilinx) evaluation board hosting the VU37P, an FPGA which includes 8GB of high bandwidth memory (HBM). In this and alos in terms of logic fabric resources this chip is more powerful than needed.  A custom production integrated board containing both four ADCs and the VU13P was chosen for the final design to improve reliability and robustness. The left panel of Figure~\ref{fig:DBE} shows the first revision of the integrated custom board.

Using the evaluation platform, the core functionality of the firmware has been completed and tested.  The fully designed and tested FPGA logic includes the following functional components:

\begin{itemize}
    \item ADC interface module 
    \item requantization block from 4-bits to 2-bits
    \item packetization module
    \item 100\,Gb/s Ethernet transmission module
    \item Universal Asynchronous Receiver Transmitter (UART)
    \item timing module
    \item monitor and control module
\end{itemize}

Additional firmware features still under development for the ngDBE include polyphase filterbank (PFB) channelization, digital slope and ripple equalization, and 1\,Gb/s Ethernet monitor and control.   The PFB and equalization features are most likely descoped from the BHEX DBE in the interests of simplicity and reduced power consumption.

The functional verification of the ngDBE firmware uses Internal Logic Analyzers (ILA) instantiated on the FPGA to look at snippets of digitized, requantized and packetized data and a recording packet capture device that ingested the output from the DBE at half rate and the data was subsequently analyzed. The integrity of a sine wave input was verified. Correlation of noise like input data generated by an antenna emulator was also successfully performed using collected data from two of the four channels.  The right panel of Figure~\ref{fig:DBE} depicts the tone verification both on the ILA and from post-processing of data captured by the external hard drive.

\section{BHEX Back End Electronics}

\label{sec:DBE}

The terrestrial progenitor systems discussed in section \ref{sec:previous_work} verify functional performance and provide firmware readily adaptable to the BHEX system.  Howeevr the hardware is not designed for flight in respect to radiation tolerance, vibration and thermal cooling in vacuum, size, weight,  and power consumption---commonly referred to as ``SWaP'' for space payloads.  The functional demonstration of a terrestrial system brings the back end, approximately,  to Technology Readiness Level 4 (TRL-4).  Using a terrestrial progenitor as an example, the R2DBE  consumes ${\sim}$75\,W, is housed in a $17''\times 18''\times 1.75''$ box, and uses key components (ADCs and FPGAs) that do not have corresponding space qualified versions.  Conversely, we aim to process the full BHEX data rate on a single, lower power digital back end (DBE) circuit board that is space qualified.
 
A Nyquist-sampled version of the received electric field in each band and both polarizations corresponds to an aggregate data rate of 64\,gigabits-per-second (Gb/s), a formidably high data rate stream to downlink from a MEO orbit to earth. No data compression on orbit is possible because the digitized signal is random noise, most of which is introduced by the receiving system, with the faint black hole emission representing only one part in 100,000 of the total power.  

VLBI detections involve cross-correlating across the ``baselines'' which connect stations---the difficult case being BHEX to any terrestrial station---followed by typically several minutes of integration. A correlation peak shows that the cosmic signal from the black hole has been detected at both stations given that the receiver noise is uncorrelated.  The cross-correlation and attendant data reduction can only be done by a processor which has access to the signal from both stations. Thus correlation cannot be done on orbit prior to downlink.  BHEX therefore requires a space-ready signal processing and transmission system that can transform the wideband analog signal to a very high rate digital data stream passing it to the optical downlink subsystem \cite{BHEX_Wang_2024} which transmits the data payload to a ground station.

\subsection{Trade Studies}

This section describes selected examples of key design trades.  Space qualified candidate components include three space qualified 10\,GS/s ADC. The AMD/Xilinx Versal FPGA is available as a space qualified part in some variants. Given our data rates and need for fast and late generation SERDES, combined with the need for lower power,  this device based on 7\,nm finFET process is a clear choice.   We are considering a ``System in Package'' (SiP) device from Mercury Systems, which uses the Versal, however this device is under Non Disclosure Agreement (NDA) and is not further described.  The possibility of using a turnkey SiP is attractive, though has cumbersome aspects including the need for NDAs, and lack of clarity and process in respect of space qualification. The Mercury SiP highlighted a possible fourth ADC candidate from Jariet, though this option is also encumbered by NDA.  An NDA is not yet in place, and so far we have not studied it.

\subsubsection{Component Trade Example: ADC Selection}
\label{sec:trade}
In this trade study  three  space qualified ADC candidates not under NDA are considered. All have maximum conversion rates $\sim$10\,GS/s.  The devices are Teledyne/e2v (EV10AS940), Analog Devices (AD9213S), and Texas Instruments (TI TI-ADC12DJ5200-SP), and their characteristics are summarized in table \ref{tab:adctrades}.   All have substantial bit depth between 8 and 10~bits.  They vary in total power consumption and analog input bandwidth. The Teledyne device having both the lowest power consumption and by far the widest analog input bandwidth of 33~GHz. The wide analog input bandwidth eliminates the need for a heterodyne stage (mixers, local oscillator, amplifiers) prior to the ADC, which is an attractive simplification of the analog preconditioning. 

However the TI ADC  provides the ability to reduce the number of lanes to four while still providing a viable and robust asynchronous JESD interface between the ADC and the FPGA.  The FPGA power consumption is driven by the number of instantiated SERDES subsystems on the FPGA. Further, for an ADC requiring as many lanes as bits, the number of  SERDES required for ADC interfacing alone, combined with the need for additional SERDES to implement fast Ethernet output drives to a multiple FPGA solution for the DBE alone.  With yet another FPGA for the optical downlink subsystem, the configuration is non-optimal in terms of power consumption, complexity, and reliability. 

Minimizing required SERDES therefore reduction of the number of  FPGAs to a single unit to outweighs the need for a heterodyne mixer stage as part of the analog conditioning.  On this basis the TI-ADC12DJ5200-SP has been provisionally selected, and the BHEX  back end presented in this paper incorporates this device specifications in the design for the back end. There are other devices still under consideration, including the SiP product mentioned earlier from Mercury Systems and the Jariet ADC.  Selection of one of these is still possible, and would then most likely be driven by considerations of simplifying some aspect of the processing elements or significant power savings.

   \begin{table}[ht]
\caption{Example of a component trade space:  Comparison of key Analog-to-Digital Converter (ADC) specifications.  All devices are capable of meeting the required conversion rate of 8.192~Gsps.  Analog input bandwidth is still a salient basis for comparison as a wide bandwidth allows for operation in 2nd or 3rd Nyquist zones,thereby needing no prior downconversion of part of the IF.} 
\label{tab:adctrades}
\begin{center}       
\begin{tabular}{|l|l|l|l|l|l|}
\hline
\rule[-1ex]{0pt}{3.5ex}  \textbf{manuf.} &\textbf{device} & \textbf{analog bw} & \textbf{power} & \textbf{\# bits} & \textbf{\# lanes}  \\
\hline
\rule[-1ex]{0pt}{3.5ex}  Teledyne e2v & EV10AS940  & 33~GHz & 2.5 W & 10 & 11 \\
\hline
\rule[-1ex]{0pt}{3.5ex}  Analog Dev. & AD9213S  & 6.5~GHz & 4.6 W & 8 & 8 \\
\hline
\rule[-1ex]{0pt}{3.5ex}  Texas Inst. &TI-ADC12DJ5200-SP   & 8~GHz & 4 W & 8 & 4 \\
\hline
\end{tabular}
\end{center}
\end{table}

\subsubsection{FPGA Count - Development Complexity Trade}

By reducing the required number of SERDES needed, the TI ADC enables a single FPGA solution for the DBE.  Even more, there is then potential to share the FPGA between the DBE and optical downlink subsystem.  Here the primary trade is related to the reduced component count, and consequently lower power and improved reliability, of the single FPGA solution.  This is traded against the relative complexity of providing appropriate interface specifications between different firmware functions, developed by different teams,  and with different legacy code-bases, within a single FPGA. These logic units  use different clock rates requiring multiple clock domains within a single FPGA.  

Layered on to this are the institutional aspects of coordinating code development between Smithsonian Astrophysical Observatory (SAO) and Lincoln Laboratories (LL), including differing repository systems, and limitations  on information sharing.  In the end, given SAO and LL are neighbors, and noting the need for close working relationships,  the significant benefits of the single FPGA solution outweigh these concerns.

\subsection{Requirements and Interface Specifications}
Following system engineering formalism considerable effort has been expended to finalize level 4 (L4), level 5 (L5) requirements, and  Interface Control Documents (ICD) for the back end. The L4 requirements focus on top level back end specifications as dictated by the science traceability matrix (STM)\cite{BHEX_Johnson_2024}. 

The L4 requirements include  sampler bandwidths, power and bandpass requirements for the analog signal before digitization, data throughput rates from the back end, data formats which are interoperable with the ground stations, and  mechanical, electrical and environmental specifications. The L5 requirements are component level specifications for the individual sub-components of the back end that derive from the L4 requirements themselves. They include hardware specifications for the converters, the FPGA, and additionally several requirements for the firmware. 

\subsection{Signal Layout}
\begin{figure*}[h]
    \centering
    \includegraphics[width=\textwidth]{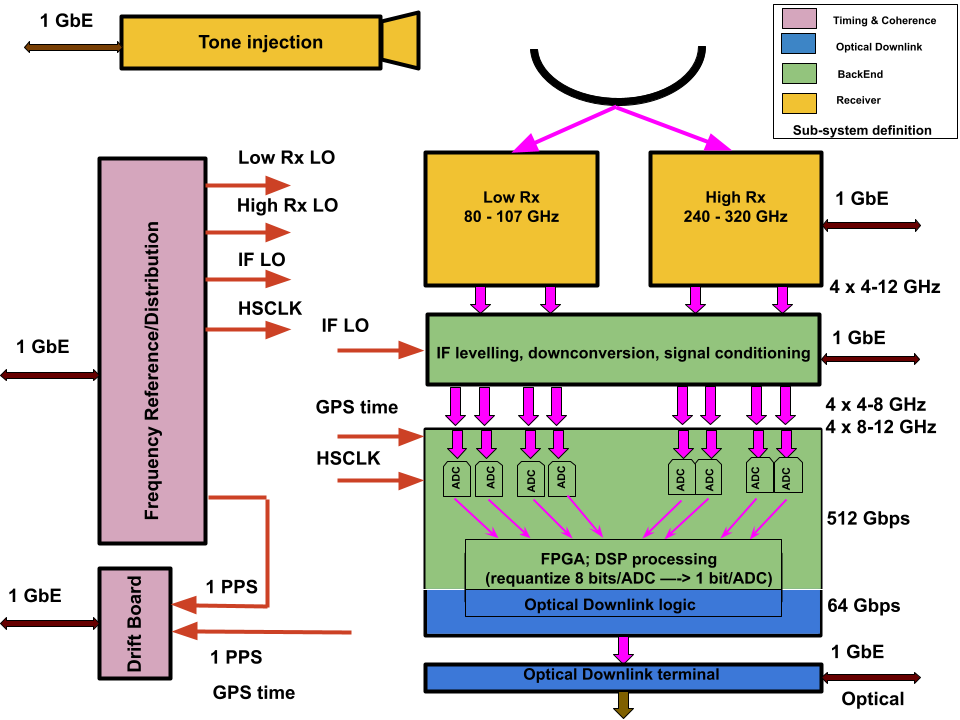}\vspace{0cm}
    \caption{System block diagram for the BHEX back end showing  interfacing sub-systems for context. From the top is shown the dish reflector, low and high band receivers, the IF pre-processing, the digital back end showing analog-to-digital converters feeding FPGA, and 64 gigabit-per-second data stream output to the optical downlink.  The back end components are color coded in green, receivers in yellow, optical downlink in blue, and frequency reference in mauve.  A single FPGA is shared between the back end and the optical downlink sub-systems to implement the firmware for both them. The single FPGA is on the DBE circuit board, and implements the optical downlink logic.  The physical layer of the optical downlink is called the optical downlink "terminal" (designated MAScOT in  reference \cite{BHEX_Wang_2024}).  The blue color code straddles two blocks in this functional block diagram to illustrate the sharing of FPGA logic resources between DBE and optical downlink with the data link between DBE and optical downlink logic implemented on-FPGA. }
\label{fig:Top level System} 
\end{figure*}

\begin{figure*}[h]
    \centering
    \includegraphics[width=\textwidth]{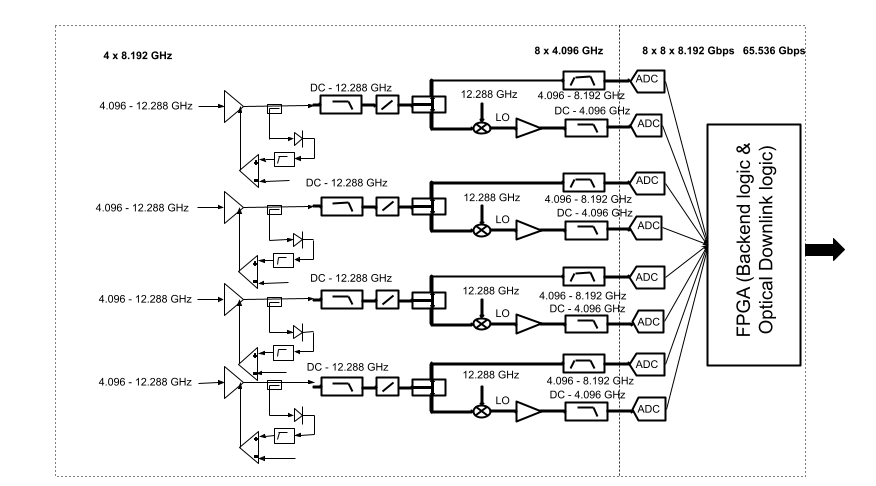}\vspace{0cm}
    \caption{ This figure breaks down the back end electronics into its constituent analog and digital sub-component elements. This  reflects a more detailed design than the system level outline in figure \ref{fig:Top level System}, including the exact frequencies of local oscillators and filters. The hashed lines reflect the most likely physical division of the analog and the digital functionality of the back end. As described in the caption to figure \ref{fig:FullBackend} the optical downlink firmware is integrated into the single FPGA represented rightmost in this figure.}
\label{fig:FullBackend} 
\end{figure*}

The cryogenic receivers operating at 1mm and 3mm output 4 IFs, each approximately 8 GHz wide which are DSB and single sideband respectively and dual polarization from 4-12 GHz. The reception and generation of the signals propagating through the entire instrument signal chain and the back end electronics is shown in the block diagram, Figure~\ref{fig:Top level System}. A more detailed schematic of the back end electronics sub-system is shown in Figure~\ref{fig:FullBackend}. This preliminary design is the most feasible option at this point given  considerations of requirements, performance, cost and space qualification. We expect it to undergo small revisions as the design matures, however we do not expect any substantial changes.

Using round numbers for convenience where appropriate, the lower sub-band is referred to as 4-8 GHz, the upper sub-band as 8-12 GHz, the total IF range as 4-12 GHz, and the instantaneous full single receive band as 8 GHz.  Where needed these are expanding to the exact frequencies documented in figure \ref{fig:FullBackend}.

\subsection{BHEX BDC}
The BHEX analog signal processing system includes a heterodyne system that performs a frequency translation from IF to baseband for parts of the band, and conditions it for digitization. The DBE is a fast digitizer sampling at 8.192 Gigasamples per second (Gsps), interfaced to digital logic implemented on a Field Programmable Gate Array (FPGA). 1-bit data output rates for each 16 GHz band with Nyquist sampling corresponds to 32 Gbps for a single frequency, or  64 Gbps for two simultaneous bands.

The   analog signal processing stage  breaks up the 8~GHz IF into two equally wide 4 GHz wide sub-bands for digitization by the subsequent stage and conditions it. The very first stage in this signal chain is the Automatic Gain Control (AGC) section which levels the amplitude to a predetermined value to keep the downstream signal chain unaffected by receiver output power variations. A Nyquist prefilter prevents  aliasing, and slope equalization corrects for gain variations across the passband These features are implemented for all IF outputs following the AGC. Two sub-bands are generated from the contiguous IF of 4-12~GHz using band pass and low pass filters. The 4-8 GHz sub-band can be direct RF sampled. The 8-12~GHz cannot be direct RF sampled and is therefore downconverted to DC-4 GHz. The depicted  configuration also introduces a spectral inversion. The mixing process causes requiring the addition of a gain  to the heterodyne system.

\subsection{BHEX DBE}
The conditioned sub-bands are digitized by eight analog to digital converters which are sampled at 8.192 Gsps.  This specific number derives from an inter-operability requirement with the EHT array. The native resolution of the TI ADC is 12-bits, however it is configurable to sample with 8-bits which  is more than adequate for our purposes. The 4-8 GHz sub-band is direct RF sampled in the second Nyquist zone of the sampler. The 8-12 GHz sub-band which is now the downconverted DC-4 GHz sub-band is sampled in the first Nyquist zone. The proposed analog to digital converter for BHEx has been identified after a component trade study as described in section \ref{sec:trade}. This has been determined to be the most optimal converter in terms of a combination of interrelated factors, including conversion rate, input bandwidth, high-rate data transfer options, power dissipation, and availability in a spaceflight version (radiation hardness, hermetic package, etc.). The trade studies for ADC's also included pushing the State of the Art, by including newly-developed microcircuits. 

\subsection{FPGA Digital Signal Processing (DSP)}
The digitized data from all the samplers are routed to an AMD Versal AI core VC1902 FPGA which is the most suitable FPGA for this application. A trade study to finalize the choice of the critical FPGA of the back end. As with the choice of converters, this FPGA is available in a spaceflight version which makes it eminently suitable. The The FPGA firmware consists of the following primary digital signal processing modules as listed and described below:
\begin{enumerate}
    \item The converter interfaces with the FPGA using a standardized multigigabit serial data standard called JESD204. This protocol is developing as the interface of choice with the selected converter employing the third revision of the protocol named JESD204C. 
    \item The 8-bit data is re-quantized down to 1 bit to reduce the data rates from a staggering 512 Gbps to a manageable 64~Gbps.
    \item The data are converted into UDP packets and relevant header information is attached to the packets in the standard VDIF format which is an EHT requirement.
    \item The UDP packets are handed off to the next sub-system over the predetermined interface.
\end{enumerate}

The ngDBE FPGA firmware described in section \ref{sec:ngdbefw} has been developed to an advanced level and has many of the features required by the above BHEX firmware. Porting from the Ultrascale+ FPGA used in ngDBE to the Versal FPGA used in BHEX is expected to be fairly straightforward with some adaptations to hardware environment needed.  It is likely that BHEX will use a stripped down feature set.  For example BHEX will only transmit one bit samples to reduce the size of data payloads on the optical downlink. Other simplifications aim to reduce firmware complexity and FPGA power consumption.  Simplifications under consideration include the elimination of polyphase filter bank (PFB) channelization and digital equalization, though the final decision on these trades is expected to be made with consideration of quantitative estimates or measurements of the actual power consumption and attendant savings. 

\subsection{Buffer Memory}
With the 16~GB as a placeholder, installing smaller, or even larger DRAM doesn’t have to change the site or package size.   System considerations including the number of terrestrial optical downlink terminals and a tolerable bit error level may mean that no  no buffer at all is needed.  The buffer memory site on the ``terrestrial equivalent'' TRL advancement hardware might then later be eliminated in a flight qualified version.  

``High Bandwidth Memory'' (HBM) is an on-chip option on modern FPGAs, available in Ultrascale+ and Versal, and can have substantial depth.  Certainly there are 8 GB versions available. However HBM is not an option in flight qualified Versal FPGAs. So there is no possibility to provide the buffer on-FPGA, and therefore the buffer size is not going to drive the 1 vs. 2 FPGA decision.   Combining the DBE and downlink Versal and firmware is a likely outcome.

\subsection{Power Estimation}
The known power consumption of the ngDBE firmware is a conservative active power estimate.  Given the  improved smaller feature size FinFET technology of Versal relative to Ultrascale+  (7~nm and 16~nm respectively) such an estimate is sure to be conservative. DBE power is a relatively small contributor to overall mission power requirements---compared to cryogenics, for example---so small variations in our estimate are of little consequence.

\section{Summary and Path Forward}
 This paper has described the BHEX system in the context of existing terrestrial VLBI systems, and  pre-design studies of the BHEX system requirements and component options. Following system engineering process, requirements tables and interface specifications to upstream and downstream systems have been developed. Top level design of the electronics hardware is described with top level, logic,  and schematic diagrams. Selection trade studies for main component parts are nearing completion, and have identified analog-to-digital converters from Texas Instruments, and the preferred FPGA in the Versal family from AMD.  
 
 The next stage is detailed design and construction of a TRL-advancement prototype of the BHEX BDC and DBE  using terrestrial grade parts which are available in functionally identical space grade equivalents.  Existing FPGA firmware developed for the terrestrial ngDBE will be adapted for BHEX and tested for fault tolerance.  Testing of the terrestrial equivalent prototype with other subsystem elements---receiver upstream, optical downlink downstream, and under coherent frequency reference discipline---aims to verify functionality. Combining careful selection of parts available in space grade versions with planned environmental testing at the University of Arizona in late 2024 aims to retire risk and further advance the BHEX back end TRL.  The objective is to prepare for a Small Explorer (SMEX) mission proposal in 2025.

\acknowledgments 
Technical and concept studies for BHEX have been supported by the Smithsonian Astrophysical Observatory, by the Internal Research and Development (IRAD) program at NASA Goddard Space Flight Center, and by the University of Arizona. We acknowledge financial support from the Brinson Foundation, the Gordon and Betty Moore Foundation (GBMF-10423), the National Science Foundation (AST-2307887, AST-1935980, and AST-2034306). This project/publication is funded in part by the Gordon and Betty Moore Foundation (Grant \#8273.01). It was also made possible through the support of a grant from the John Templeton Foundation (Grant \#62286).  The opinions expressed in this publication are those of the author(s) and do not necessarily reflect the views of these Foundations. BHEX is funded in part by generous support from Mr. Michael Tuteur and Amy Tuteur, MD. BHEX is supported by initial funding from Fred Ehrsam.

\end{document}